\documentclass[%
superscriptaddress,
twocolumn,
aps,
prl,
]{revtex4-1}

\usepackage{graphicx}
\usepackage{dcolumn}
\usepackage{bm}
\usepackage{amsmath,amsthm,amssymb}
\usepackage[separate-uncertainty = true]{siunitx} 
\usepackage{color}
\DeclareSIUnit\permille{\text{\textperthousand}}
\usepackage[english]{babel}
\usepackage{amsfonts}
\usepackage{appendix}
\usepackage{textcomp} 

\begin{document}

\preprint{APS/123-QED}
\title{Matched Guiding and Controlled Injection in Dark-Current-Free, 10-GeV-Class, Channel-Guided Laser Plasma Accelerators \\
}

\author{A. Picksley}
\email{apicksley@lbl.gov}
\affiliation{Lawrence Berkeley National Laboratory, Berkeley, California 94720, USA}%
\author{J. Stackhouse}
\affiliation{Lawrence Berkeley National Laboratory, Berkeley, California 94720, USA}%
\affiliation{University of California, Berkeley, California 94720, USA}%
\author{C. Benedetti}
\affiliation{Lawrence Berkeley National Laboratory, Berkeley, California 94720, USA}%
\author{K. Nakamura}
\affiliation{Lawrence Berkeley National Laboratory, Berkeley, California 94720, USA}%
\author{H. E. Tsai}
\affiliation{Lawrence Berkeley National Laboratory, Berkeley, California 94720, USA}%
\author{R. Li}
\affiliation{Lawrence Berkeley National Laboratory, Berkeley, California 94720, USA}%
\affiliation{University of California, Berkeley, California 94720, USA}%
\author{B. Miao}
\affiliation{Institute of Research in Electronics and Applied Physics and Dept. of Physics, University of Maryland, College Park, MD 20742, USA}%
\author{J. E. Shrock}
\affiliation{Institute of Research in Electronics and Applied Physics and Dept. of Physics, University of Maryland, College Park, MD 20742, USA}%
\author{E. Rockafellow}
\affiliation{Institute of Research in Electronics and Applied Physics and Dept. of Physics, University of Maryland, College Park, MD 20742, USA}%
\author{H. M. Milchberg}
\affiliation{Institute of Research in Electronics and Applied Physics and Dept. of Physics, University of Maryland, College Park, MD 20742, USA}
\affiliation{Dept. of Electrical and Computer Engineering, University of Maryland, College Park, MD 20742, USA}
\author{C. B. Schroeder}
\affiliation{Lawrence Berkeley National Laboratory, Berkeley, California 94720, USA}%
\affiliation{University of California, Berkeley, California 94720, USA}%
\author{J. van Tilborg}
\affiliation{Lawrence Berkeley National Laboratory, Berkeley, California 94720, USA}%
\author{E. Esarey}
\affiliation{Lawrence Berkeley National Laboratory, Berkeley, California 94720, USA}%
\author{C. G. R. Geddes}
\affiliation{Lawrence Berkeley National Laboratory, Berkeley, California 94720, USA}%
\author{A. J. Gonsalves}
\affiliation{Lawrence Berkeley National Laboratory, Berkeley, California 94720, USA}%

\date{\today}

\begin{abstract}
We measure the high intensity laser propagation throughout meter-scale, channel-guided LPAs by adjusting the length of the plasma channel on a shot-by-shot basis, showing high quality guiding of $\SI{500}{TW}$ laser pulses over 30 cm in a hydrogen plasma of density $n_0 \approx \SI{1e17}{cm^{-3}}$.
We observed transverse energy transport of higher-order modes in the first $\approx \SI{12}{cm}$ of the plasma channel, followed by quasi-matched propagation, and the gradual, dark-current-free depletion of laser energy to the wakefield. We quantify the laser-to-wake transfer efficiency limitations of currently available PW-class laser systems, and demonstrate via simulation how control over the laser mode can significantly improve accelerated beam parameters. Using just $\SI{21.3}{J}$ of laser energy, and triggering localized electron injection into the accelerator, we observed electron bunches with single, quasimonoenergetic peaks, relative energy spreads as low as $\SI{3}{\%}$ and energy up to $\SI{9.2}{GeV}$ with charge extending beyond $\SI{10}{GeV}$.

\end{abstract}

\maketitle

Acceleration of particles in plasma waves driven by intense laser pulses \cite{Dawson1979, Esarey.2009, Hooker2013} has attracted significant attention due to the ultra-high accelerating gradients that can be achieved (up to 100s GV/m). These compact accelerators are attractive for applications such as free-electron lasers \cite{Wang2021, Labat2023}, Thomson sources \cite{Powers.2013}, and high-energy physics colliders \cite{Schroeder2010, schroeder2023linear}. Significant progress has been made towards producing high-energy ($> \SI{1}{GeV}$) \cite{Leemans2006, Clayton2010, Lu2011, Wang2013, Leemans2014, Gonsalves2019, Miao2022, aniculaesei2024acceleration}, high-quality \cite{Faure2006, Lundh2011, Plateau2012, Wang2016} electron bunches in laser-plasma accelerators (LPAs). 

Maximizing electron beam energy for a given laser energy is critical for these applications. This requires maintaining laser intensity over several tens of centimeters, much longer than the typical Rayleigh length $z_\mathrm{R} \sim \SI{1}{cm}$ of focused PW-class laser systems. Hence, laser pulses must be guided via relativistic self-focusing \cite{Sprangle1990} or using a preformed plasma channel \cite{sprangle1992interaction, Durfee1993}. The former requires operating in the bubble regime of the LPA \cite{Pukhov2002, Lu2006}, at a plasma density $n_0$ high enough to remain above the critical power for self-focusing. High-energy electrons can be produced but high laser pulse energy is required \cite{Esarey.2009}; for example Aniculaesei \textit{et al.} \cite{aniculaesei2024acceleration} demonstrated an energy gain of $\sim \SI{10}{GeV}$ using $\SI{118}{J}$ of laser energy. Preformed plasma channels have transverse electron density profiles with a minimum on-axis such that the refractive index is peaked, much like gradient-index optical fibers. They allow operation of LPAs at lower $n_0$ where the product of acceleration gradient and accelerator length is higher.

Matching the drive laser to the fundamental transverse mode of the plasma channel maximises the efficiency of laser-to-wake energy transfer without degrading the accelerated bunch parameters, and minimizes the accelerator dark current. For a parabolic plasma channel where $n_\mathrm{e}(r)-n_0 \propto r^2$, a low-intensity laser pulse with a Gaussian transverse profile is perfectly matched to the fundamental channel mode and propagates at constant spot-size if $w_0 = w_\mathrm{m}$, where $w_\mathrm{m}$ is the matched spot-size, a measure of the steepness of the parabolic profile \cite{sprangle1992interaction, Esarey.2009}. Future applications of 10-GeV-class LPAs require matched propagation of PW-class lasers with $w_\mathrm{m} \lesssim \SI{50}{\micro m}$ at $n_0 \approx \SI{1.0e17}{cm^{-3}}$ \cite{Schroeder2010, schroeder2023linear}. 

Capillary-discharge waveguides \cite{Butler2002, Spence2003} have previously been employed in multi-GeV LPAs \cite{Leemans2006, Leemans2014}. Whilst they are approximately parabolic, and can provide per-mille level stability of $w_\mathrm{m}$ and $n_0$ \cite{turner2021radial}, $w_\mathrm{m}$ is too large for sufficient confinement at $n_0 \approx \SI{e17}{cm^{-3}}$. Even with the addition of an auxillary laser to reduce $w_\mathrm{m}$ \cite{Bobrova2013, pieronek2020laser}, effective guiding was only achieved above optimal $n_0$, limiting energy gain to $\SI{7.8}{GeV}$ \cite{Gonsalves2019}.

Building upon the previous demonstrations of hydrodynamically formed channels that employed collisional heating \cite{Durfee1993, Volfbeyn1999} and were limited to $n_0 \sim \SI{e19}{cm^{-3}}$, laser heating with $\lesssim \SI{100}{fs}$ pulses has been employed \cite{Lemos2013, Lemos2013a, Shalloo2018, Shalloo2019}. The heating mechanism, optical field ionization, is effective at lower density \cite{Shalloo2018, Shalloo2019}. A cylindrical column of plasma is formed in the line focus of an ultrashort laser pulse, and expands radially. When an intense drive laser is focused into the plasma channel, the leading edge ionizes the neutral gas surrounding the expanding shock front to create a deep, thick-walled plasma channel \cite{robert2018a, Morozov2018, Picksley2020a, Feder2020}. Steep channels formed by hydrodynamic expansion of optical field-ionized (HOFI) plasmas ($\SI{20}{\micro m} \lesssim w_\mathrm{m} \lesssim \SI{50}{\micro m}$) with $n_0 \lesssim \SI{e17}{cm^{-3}}$ can be generated \cite{Shalloo2018, robert2018a, Picksley2020a, Miao2020, Feder2020, Mewes2023}. Guiding of high intensity pulses \cite{Shalloo2019, Smartsev2019, Picksley2020, Picksley2020a, Feder2020, Ross2024}, and electron acceleration \cite{Oubrerie2022, Miao2022, Picksley2023} have been previously demonstrated, with $\sim \SI{5}{GeV}$ the highest energy to date \cite{Miao2022}. In such experiments, guiding is assessed at the end of the accelerator, and propagation of the laser pulse throughout the plasma is inferred from simulations. Laser mode beating and evolution theory \cite{Esarey1999, Esarey2000, Benedetti2012, Benedetti2015} was recently analyzed by Shrock \textit{et al.} \cite{Shrock2024} via PIC simulations showing mismatched laser guiding linked to measured electron spectra.

In this Letter, we measure high-intensity laser propagation throughout meter-scale LPAs by adjusting the length of the accelerator on a shot-by-shot basis for the first time, showing high quality guiding of $\SI{500}{TW}$ laser pulses throughout a 30-cm-long hydrogen plasma of density $n_0 \approx \SI{1e17}{cm^{-3}}$. Using extensive optical diagnostics, we show laser pulse coupling into the higher-order channel modes and energy loss through mode-filtering, followed by quasi-matched propagation of the fundamental mode, and the gradual, dark-current-free depletion of laser energy to the plasma wave. Then, by triggering electron injection via the localized addition of nitrogen to the plasma, electron bunches with single, quasi-monoenergetic peaks up to $\SI{9.2}{GeV}$ and charge extending to $> \SI{10}{GeV}$ is achieved using just $\SI{21.3(3)}{J}$ of laser energy. We quantify the laser-to-wake transfer efficiency limitations of currently available PW-class laser systems, and demonstrate via simulation how control over the laser mode can result in $\gtrsim \SI{13}{GeV}$ bunches for the same plasma channel. This work provides in-depth understanding of non-linear laser evolution and electron acceleration in 10-GeV-class LPA stages suitable for future compact accelerators.

\begin{figure}[t!]
    \centering
    \includegraphics[width=\linewidth]{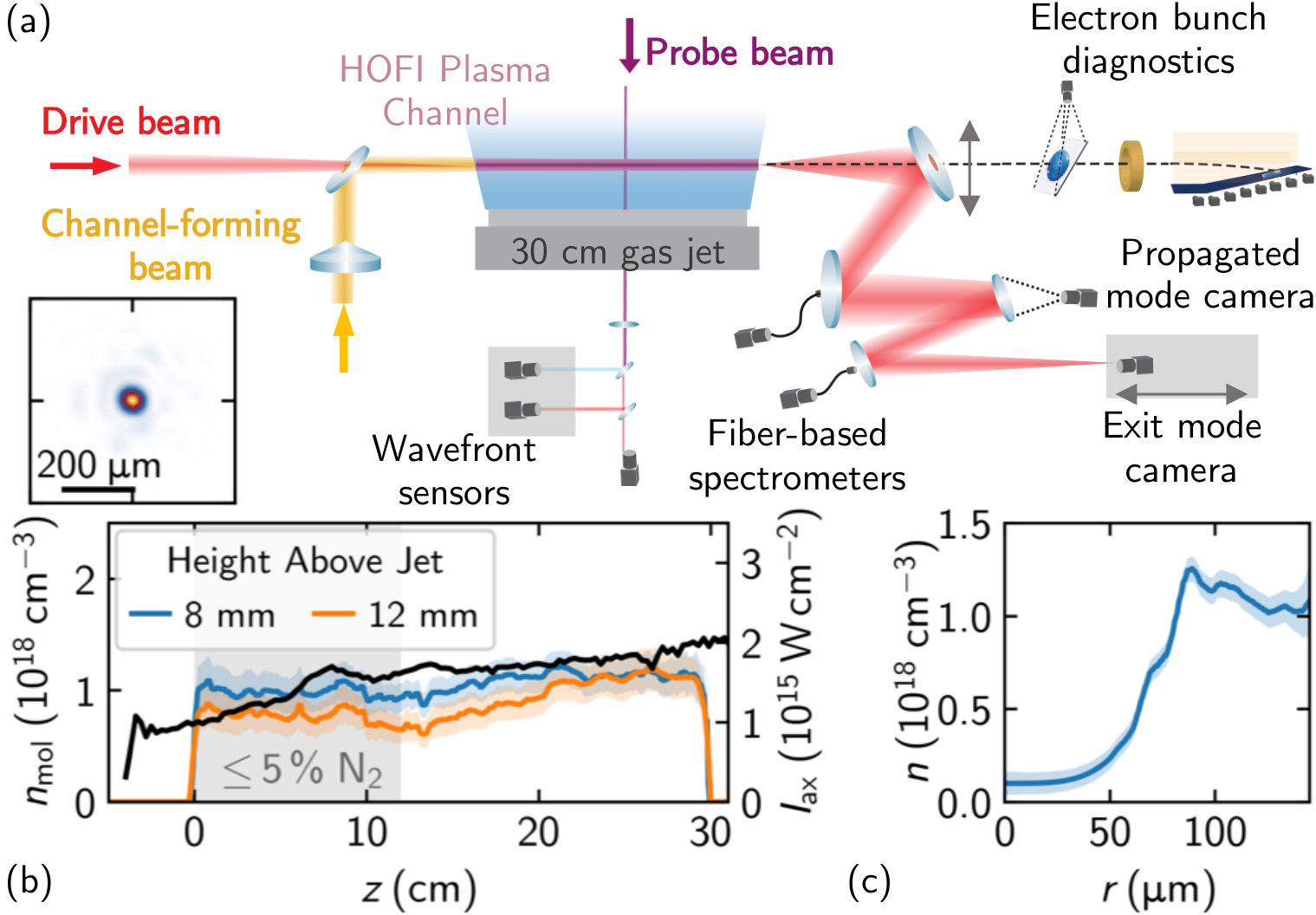}
    \caption{(a) Schematic of the experimental setup. Inset: Measured vacuum mode of the drive laser pulse. (b) Measured molecular density of the gas jet (blue and orange), and peak intensity of the channel-forming pulse along the length of the gas (black). (c) Measured electron and neutral density $n = n_\mathrm{e} + n_\mathrm{n}$ of the HOFI plasma channel at $\Delta \tau = \SI{6}{ns}$. 
    }
    \label{fig:setup}
\end{figure}

The Ti:Sapphire-based BELLA petawatt laser \cite{Nakamura2017} produces pulses of full-width-half-maximum (FWHM) duration $\sim \SI{40}{fs}$ at a central wavelength $\lambda_0 = \SI{815}{nm}$. Recent upgrades allow the amplified laser to be split into two separately compressed beamlines \cite{turner2022strong} with control over the relative timing, wavefront, and focusing geometry. Figure \ref{fig:setup}(a) shows a schematic of the experimental setup. In the channel-forming beamline, $\SI{1.3(3)}{J}$ was focused by an axicon lens and reflected by a mirror with a hole drilled in the center into the gas target \cite{Shalloo2019, Picksley2020, Picksley2020a, Miao2022}. The peak intensity $I_\mathrm{ax}$ as a function of distance from the entrance of the gas jet, $z$, is shown (black) in figure \ref{fig:setup}(b). The drive pulse was focused to a spot-size $w_0 = \SI{53(1)}{\micro m}$ at the entrance of the gas target. The energy was varied up to a maximum of $\mathcal{E}_0 = \SI{21.3(3)}{J}$, corresponding to a peak normalized vector potential $a_0 \approx 0.85 \lambda [\SI{}{\micro m}] \sqrt{I_0 [\SI{e18}{W.cm^{-2}}]}$ of up to $a_0 \approx 2.2$. Here, $I_0$ is the peak intensity of the drive pulse.

The drive laser and electron diagnostics have been described previously \cite{Nakamura2017, Gonsalves2019}. The input and guided mode of the drive laser could be imaged over a range of $\sim \SI{60}{cm}$; the propagated drive was also imaged at the plane of the third wedge $\approx \SI{10}{m}$ downstream of the channel exit. The optical spectrum was measured using fiber-based spectrometers covering the range $\SI{400}{nm} \lesssim \lambda \lesssim \SI{2200}{nm}$. The energy transmission $T(z) = \mathcal{E}(z)/\mathcal{E}_0$ was retrieved by integrating the counts on the detector at $z \approx \SI{10}{m}$, and then using the measured optical spectrum to correct for the spectral response of the detector. 

For this experiment, a 30-cm-long gas target was developed \cite{Krishnan2011, zhou2021effect, Miao2022}. The jet comprised an elongated, converging-diverging nozzle operated with hydrogen, or hydrogen with a $\leq \SI{5}{\%}$ nitrogen dopant. The length could be varied by blocking the flow of gas above the nozzle. Figure \ref{fig:setup}(b) shows the molecular density as a function of distance along the gas jet measured using method outlined in \cite{Miao2022}. The delay between the channel-forming beam and the drive beam was set to $\Delta \tau = \SI{6}{ns}$, and the jet was operated $\SI{12}{mm}$ below the laser axis to avoid blocking the channel-forming beam (which had a radius of $\approx \SI{12}{mm}$ at $z=0$). For these conditions, two-color interferometry measurements \cite{Gonsalves2007, Point2014, Feder2020} shown in figure \ref{fig:setup}(c) indicated an axial plasma density $n_0 \approx \SI{1e17}{cm^{-3}}$ and matched spot-size $w_\mathrm{m} \approx \SI{37}{\micro m}$. We note that due to the large uncertainties associated with two-color interferometry, even though the channel expanded further with increasing $\Delta \tau$, the calculated $w_\mathrm{m}$ was unchanged for delays $\SIrange{5}{7}{ns}$ used in this work.

\begin{figure}[t]
    \centering
    \includegraphics[width=\linewidth]{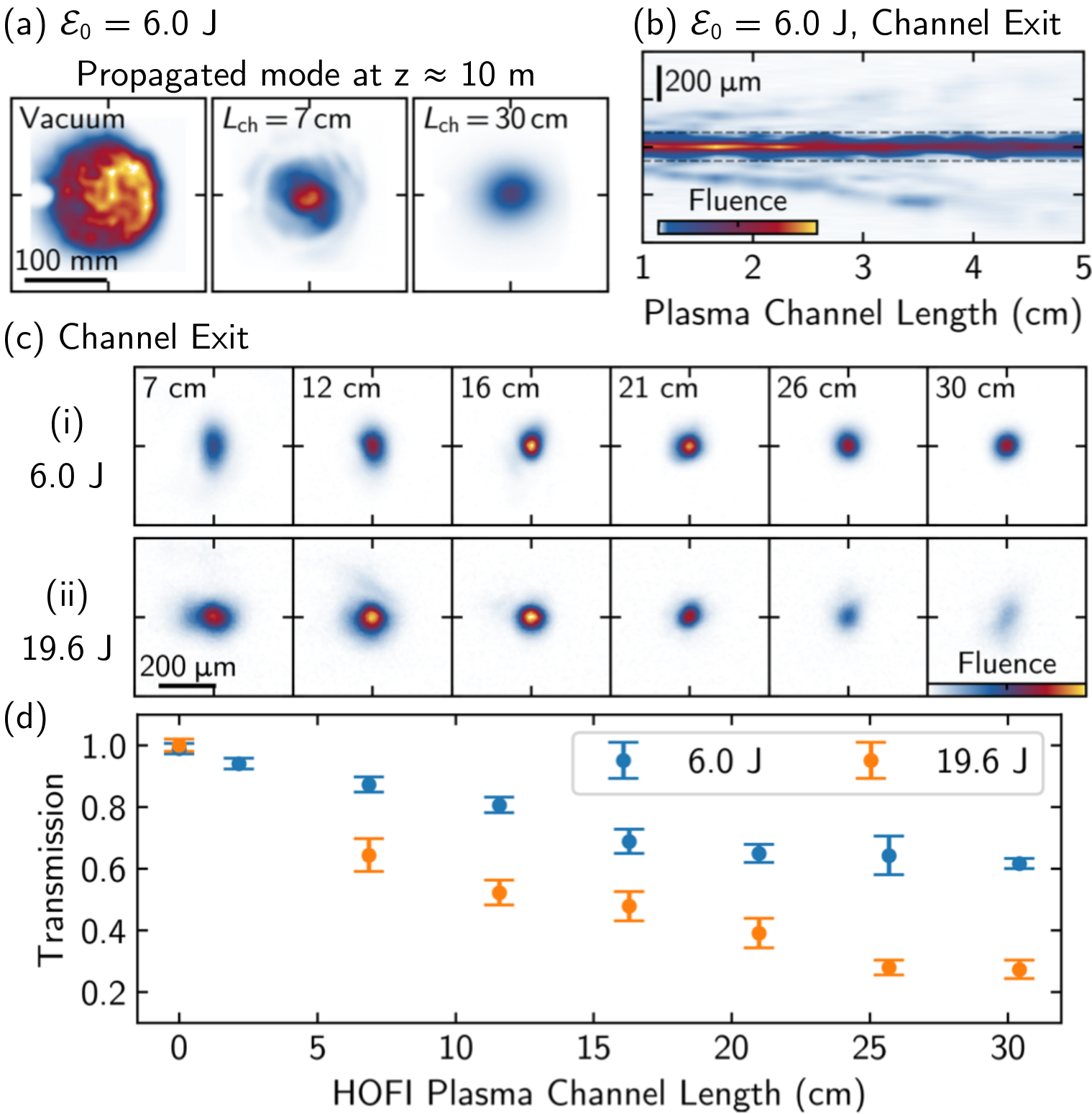}
    \caption{Evolution of the drive laser in HOFI plasma channels with $n_0 \approx \SI{1e17}{cm^{-3}}$ and $w_\mathrm{m} \approx \SI{37}{\micro m}$. (a) Propagated mode $z \approx \SI{10}{m}$ downstream. (b) Lineouts of the exit mode for $\mathcal{E}_0 = \SI{6.0(1)}{J}$ for short channel lengths. The dashed line indicates the measured peak of the neutral density profile. (c) Exit modes for $\mathcal{E}_0 = \SI{6.0(1)}{J}$ and $\SI{19.6(4)}{J}$. (d) Measured energy transmission averaged over $\approx 20$ shots. 
    }
    \label{fig:guiding}
\end{figure}

The evolution of key drive laser parameters for two different laser energies $\mathcal{E}_0 = \SI{6.0(1)}{J}$ ($a_0 \approx 1.3$) and $\SI{19.6(4)}{J}$ ($a_0 \approx 2.2$) is shown in figure \ref{fig:guiding}. Only shots for which the transverse position of the laser focus with respect to the channel entrance $\Delta R < \SI{25}{\micro m}$ were included, inferred from a non-destructive centroid diagnostic \cite{isono2021high}. This condition was satisfied for $\SI{71}{\%}$ of shots. 
Figure ~\ref{fig:guiding}(a) shows representative transverse fluence profiles of the drive laser for several different channel lengths $\approx \SI{10}{m}$ downstream of the waveguide exit. For $L_\mathrm{ch} \approx \SI{7}{cm}$ the drive laser mode had transformed from the top-hat-like input mode (that is typical of currently available PW-class systems based on bulk crystal) to near-Gaussian; the super-Gaussian \cite{superGauss} fit order reduced from $\approx 6$ to $\approx 2$. Through $z \approx \SI{12}{cm}$, the propagated mode exhibited rings outside the central fluence peak. As the channel was lengthened, a single, approximately Gaussian transverse fluence profile was always observed when the drive laser was well-aligned to the channel. Figure \ref{fig:guiding}(b) shows a waterfall plot of lineouts from the re-imaged exit mode, with the channel length varied in steps of $\SI{0.2}{cm}$ up to $L_\mathrm{ch} \approx \SI{5}{cm}$. The dashed line indicates the measured peak of $n(r)$. Figure \ref{fig:guiding}(c) shows re-imaged exit mode images for channel lengths up to $L_\mathrm{cm} \approx \SI{30}{cm}$. For $z \gtrsim \SI{12}{cm}$, a well-confined, near-Gaussian drive mode was observed for all channel lengths and for both laser intensities.

The dominant mechanisms behind laser pulse propagation \cite{Esarey1999, Esarey2000, Clark2000_mode, Shrock2024} can be understood from guiding measurements presented in figure \ref{fig:guiding}. The channel supports several quasi-bound transverse modes which we denote as $(p,m)$ referring to the radial and azimuthal mode number respectively. Their structure are determined by $n_\mathrm{e}(r)$. HOFI plasma channels are finite in extent and not radially parabolic; only a finite number of low-order modes can propagate, of which the fundamental $(0,0)$ mode is close to Gaussian. Since the input laser mode [shown in figure \ref{fig:setup}(a)] was not the fundamental channel mode, energy was coupled into higher-order modes, observed directly from the ring structure in figure \ref{fig:guiding}(a) for $L_\mathrm{ch} \approx \SI{7}{cm}$. Figure \ref{fig:guiding}(b) shows rapid leakage of higher-order modes out of the bound region at oblique angles over a few cm \cite{Clark2000_mode}. These modes do not contribute to wakefield generation. After this initial period of mode-filtering, which occurred over $z \lesssim \SI{12}{cm}$, higher-order mode content in the guided mode was severely reduced, evidenced by figure \ref{fig:guiding}(c,i). Remaining higher-order modes slip behind the fundamental due to group velocity dispersion, and eventually become separated longitudinally such that they also do not contribute to wakefield generation \cite{Esarey1999, Esarey2000, Schroeder2011, cormier2011control, van2014measurement, djordjevic2019control, djordjevic2018filtering, Shrock2024}. The distance over which the $(1,0)$ and $(2,0)$ modes separate by $\SI{40}{fs}$ was calculated to be $\SI{21.6}{cm}$, and $\SI{9.4}{cm}$ respectively. For $z \gtrsim \SI{12}{cm}$, the measured exit mode remained approximately Gaussian with measured spot-size oscillation $\lesssim 6 \%$, demonstrating approximately matched propagation of the drive in the fundamental mode.

Coupling and propagation losses were evaluated quantitatively for the laser and channel used here. The mode coupling efficiency $\eta$ of the input drive laser into the measured fundamental mode was calculated as the overlap integral between the two modes. For the measured input spot ($w_0 \approx \SI{53}{\micro m}$), $\eta \approx \SI{60}{\%}$. The focus of a perfect flat-top-like laser with the same $w_0$ yields $\eta \approx \SI{72}{\%}$, and for a size of $w_\mathrm{0} \approx \SI{47}{\micro m}$, $\eta$ can be as high as $\approx \SI{85}{\%}$. Measured propagation loss [see figure \ref{fig:guiding}(d)] occurred when light exited the channel outside the acceptance angle of our diagnostics ($\gtrsim \SI{0.75}{deg}$) or when higher-order modes that remained at the end of the channel diffracted. For the latter, it was calculated that $> \SI{99}{\%}$, $\SI{85}{\%}$ and $\SI{28}{\%}$ of light remaining in the $(0,0)$, $(1,0)$ and $(2,0)$ modes respectively was captured inside the diagnostic acceptance angle. For $\mathcal{E}_0 = \SI{6.0(1)}{J}$, measured $T(z)$ indicates $\SI{31(4)}{\%}$ losses in the first $\SI{16}{cm}$, consistent with our predicted coupling losses. Further losses for $\SI{16}{cm} \lesssim z \lesssim \SI{30}{cm}$ were $\lesssim \SI{10}{\%}$ indicating mild coupling of remaining energy to the wake. The calculated attenuation through leakage of light remaining in the $(1,0)$ and $(0,0)$ modes was small over $\SI{16}{cm} \lesssim z \lesssim \SI{30}{cm}$. Since wakefield generation is mainly driven by energy in the $(0,0)$ mode, and higher-order modes are filtered, control over the channel length directly illustrates how laser-to-wake transfer efficiency is limited in LPAs due to currently available PW-class systems, and can be maximized by careful matching of the laser size to the channel.

\begin{figure}[t!]
    \centering
    \includegraphics[width=\linewidth]{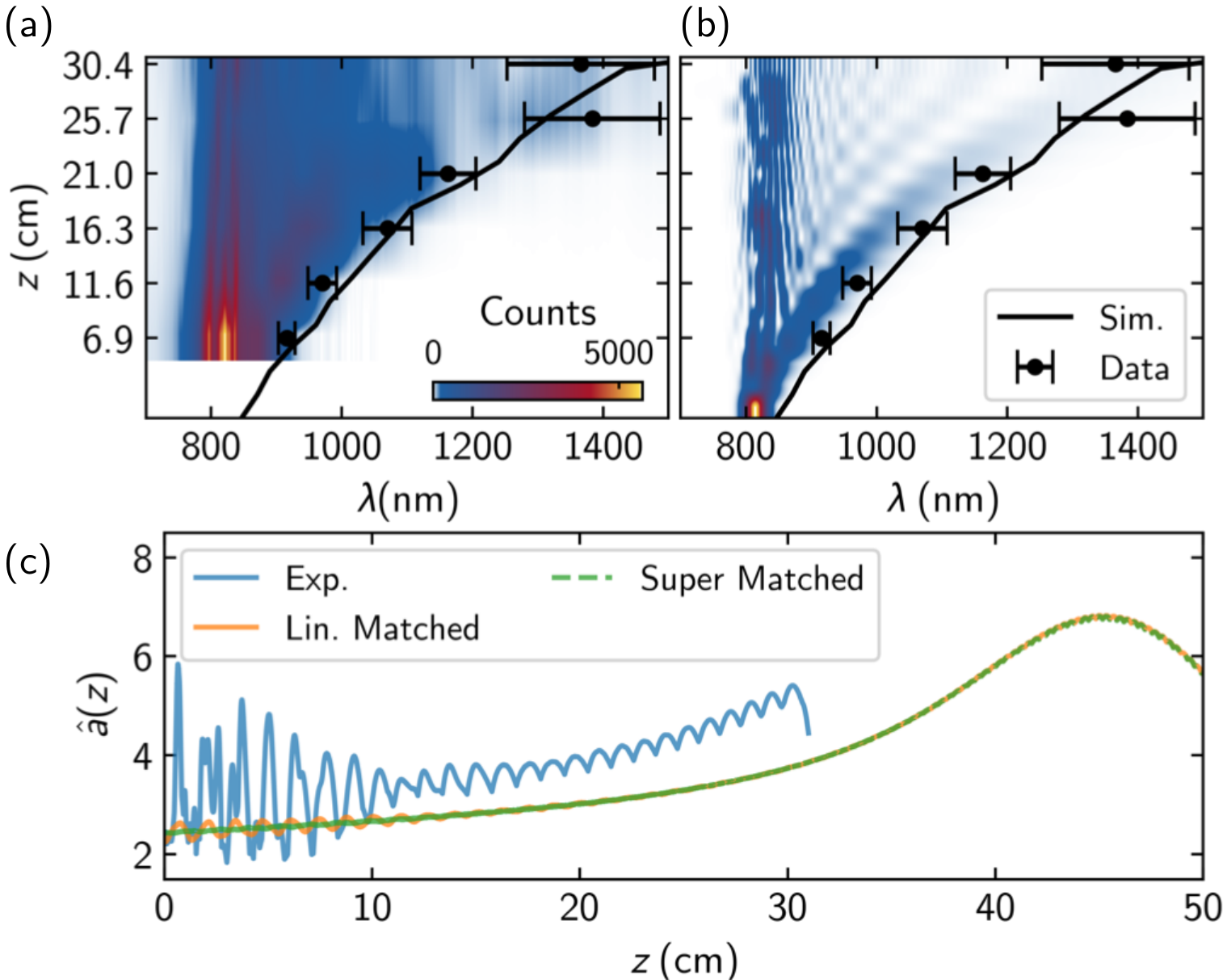}
    \caption{Measured (a) and simulated (b) optical spectra as a function of propagation distance for $\mathcal{E}_0 = \SI{19.6}{J}$, $n_0 \approx \SI{1e17}{cm^{-3}}$ and $w_\mathrm{m} \approx \SI{37}{\micro m}$. Measured (averaged over $\approx 20$ shots) and calculated $\lambda_\mathrm{R}$ is overlaid in black. (c) Calculated normalized peak laser intensity $\hat{a}$  as a function of propagation distance for the experiment input spot, linearly matched, and super-matched spots. 
    }
    \label{fig:spectrum}
\end{figure}

Driving a wakefield suitable of generating multi-GeV beams on the length scales of our gas jet required increased laser intensity \cite{Esarey.2009}. Similar behavior of the drive laser evolution for laser energy $\SI{19.6(4)}{J}$ is observed in figure \ref{fig:guiding}(b, ii), with the exception of decreased transmission associated with wakefield generation. This was confirmed by gradual laser redshifting shown in figure \ref{fig:spectrum}(a). Simulations of this case were performed using the code \textsc{INF\&RNO} \cite{benedetti2010efficient, benedetti_2018}. The measured parameters for the energy, temporal profile, and transverse profile of the drive laser (which was recovered using a Gerchberg-Saxton algorithm) were input, and $n(r)$ was set to the measured channel for $\Delta \tau$ [see figure \ref{fig:setup}(c)]. Figure \ref{fig:spectrum}(b) shows the calculated optical spectrum during propagation. The wavelength at which the spectrum reduces to $\SI{5}{\%}$ of the peak, $\lambda_\mathrm{R}$ is shown in black and matched closely to experiment for all $L_\mathrm{ch}$, explicitly demonstrating continual depletion of energy from the laser to the plasma wave. The calculated average field was $E_z \approx \SI{30}{GV.m^{-1}}$. Unlike for $\mathcal{E}_0 = \SI{6.0(1)}{J}$, at high-intensity $T(z)$ continuously reduced as energy was coupled to the wake \cite{Shadwick2009, Benedetti2015, Picksley2020a, Miao2022}. 

Figure \ref{fig:spectrum}(c, blue) shows the calculated evolution of the normalized peak laser intensity $\hat{a}(z)$ for parameters of figure \ref{fig:spectrum}(b). Oscillations in $\hat{a}(z)$ are due to mode beating and cause periodic changes in the longitudinal and transverse structure of the wakefield \cite{Esarey1999, Benedetti2012, Benedetti2015, Shrock2024}. Changes in the longitudinal structure result from relativistic effects (i.e., the dependence of the plasma wavelength on the laser peak strength), while changes in the transverse structure result from laser mode evolution (i.e., the shape of the laser mode varies because of mode beating, and this affects the transverse component of the ponderomotive force). The latter can result in a wakefield that is unsuitable for the transport of electron beams if, for instance, the laser mode acquires a sufficiently deep minimum on axis. Mode filtering and subsequent mode dispersion reduce the visibility of oscillations for $z \gtrsim \SI{12}{cm}$, consistent with the diminishing ring structures in figure \ref{fig:guiding}(a). Low visibility oscillations caused by beating between the $(0,0)$ and $(1,0)$ modes are present as the laser self-steepens and redshifts.

\begin{figure}[t!]
    \centering
    \includegraphics[width=\linewidth]{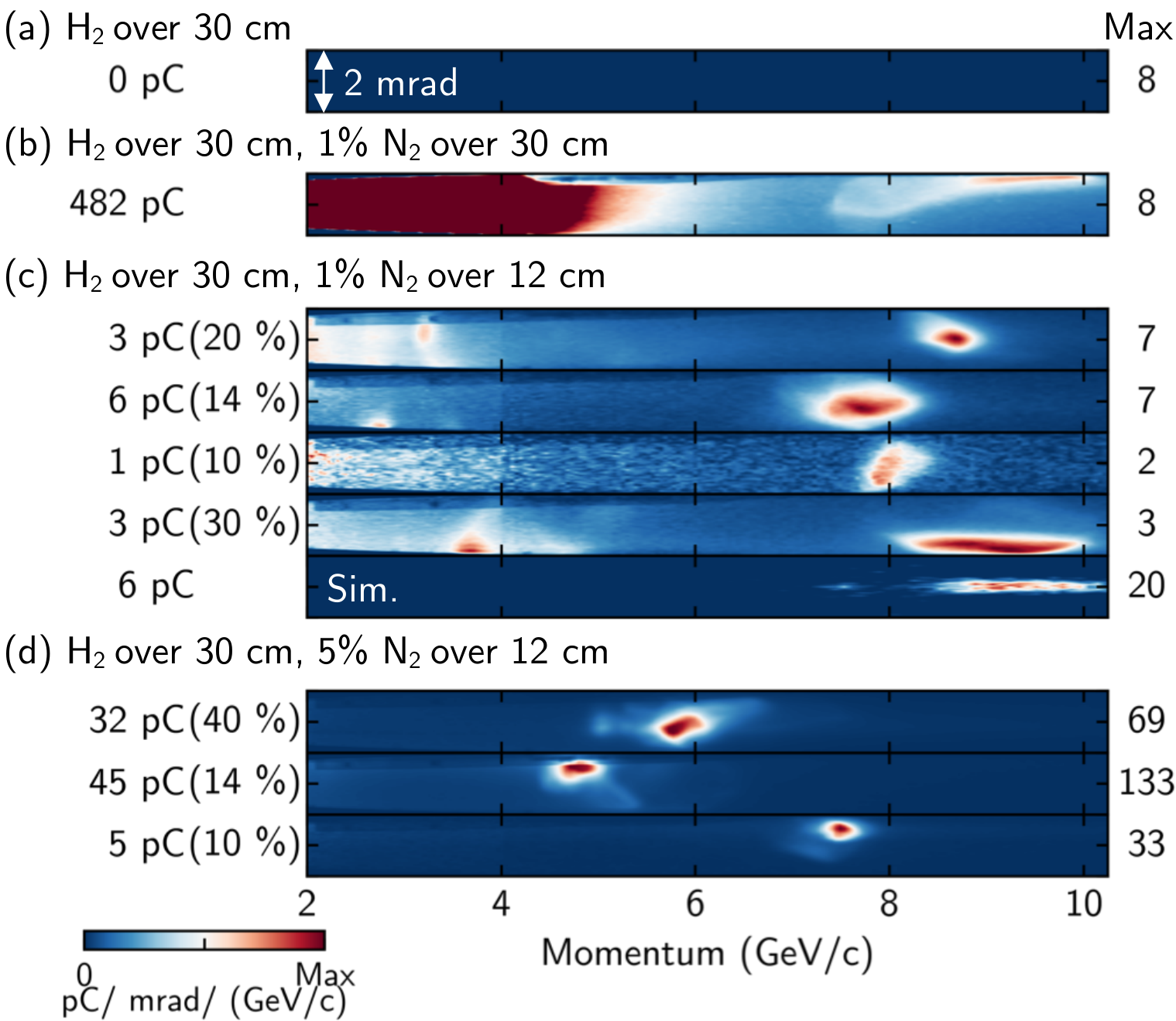}
    \caption{Example electron beams generated in 30-cm-long HOFI plasma channels with $\mathcal{E}_0 = \SI{21.3(3)}{J}$. For each row, the charge measured by the spectrometer within the quasimonoegergetic bunch and percent captured by the spectrometer is given. (a) $\Delta \tau = \SI{6}{ns}$, no nitrogen, (b) $\Delta \tau = \SI{7}{ns}$, \SI{1}{\%} nitrogen, $L_\mathrm{dop} \approx \SI{30}{cm}$, (c) $\Delta \tau = \SI{5}{ns}$, $\SI{1}{\%}$ nitrogen, $L_\mathrm{dop} \approx \SI{12}{cm}$, (d) $\Delta \tau = \SI{6}{ns}$, $\SI{5}{\%}$ nitrogen, $L_\mathrm{dop} \approx \SI{12}{cm}$. 
    }
    \label{fig:beams}
\end{figure}

No electron beams were generated for the experiments presented above [see figure \ref{fig:beams}(a)], demonstrating that laser pulses can be well-guided for densities below the self-trapping threshold at laser intensities sufficient to generate high amplitude plasma waves. Electron beams were generated by introducing a nitrogen dopant to the gas jet \cite{Pak2010, McGuffey2010, Chen2012}. A $\SI{1}{\%}$ dopant extending throughout the jet triggered injection at several points, and resulted in electron bunch spectra with a broad distribution. Electrons injected after short propagation distance experience the plasma wave over a longer distance and reach higher energies, whilst electrons injected later in the plasma channel experience less energy gain. An example bunch with conditions similar to figure \ref{fig:guiding} (but with $\Delta \tau = \SI{7}{ns}$) is shown in figure \ref{fig:beams}(b). A peak in the tail of the distribution was observed at $\sim \SI{9.4}{GeV}$ with charge extending $\gtrsim \SI{10}{GeV}$. 

To study the acceleration of single, quasimonoenergetic bunches, we restricted the dopant region $0 \leq z \lesssim L_\mathrm{dop}$ within the gas jet \cite{Gonsalves2020, Kirchen2020, Shrock2024}. For $L_\mathrm{dop} \approx \SI{6}{cm}$, high energy electrons were not observed. Figure \ref{fig:beams}(c) shows generated beams for $L_\mathrm{dop} \approx \SI{12}{cm}$, $\Delta \tau = \SI{5}{ns}$, and $\mathcal{E}_0 = \SI{21.3(3)}{J}$. Singly-peaked electron bunches were observed, indicating injection in the region $\SI{6}{cm} \lesssim z \lesssim \SI{12}{cm}$. The mean energy, and FWHM spread for the examples in figure \ref{fig:beams}(c) were $\SI{8.67(0.48)}{GeV}$, $\SI{7.70(0.88)}{GeV}$, $\SI{7.96(0.44)}{GeV}$ and $\SI{9.15(1.80)}{GeV}$. Shot-to-shot stability was dominated by transverse offset of the laser focus at the plasma channel entrance, and by variations of $\gtrsim \SI{20}{\%}$ in the pulse duration. Due to pointing variations and limited acceptance of the spectrometer, not all of the charge recorded by the phosphor screen was captured by the magnetic spectrometer. For each example, the measured charge within the quasimonoenergetic bunch and percentage of charge captured is shown. The bottom panel in figure \ref{fig:beams}(c) shows results from \textsc{INF\&RNO} simulations with the same conditions. The simulation confirmed ionization of nitrogen occurred throughout the dopant region, $z \leq \SI{12}{cm}$, however changes in the wake structure noted in figure \ref{fig:spectrum}(a) prevented the trapping of electron bunches with a significant charge for $z \lesssim \SI{8.6}{cm}$ \cite{Picksley2023, Shrock2024}. A portion of the electrons ionized within $\SI{8.6}{cm} \lesssim z \lesssim \SI{12}{cm}$ were accelerated to $\SI{9.3}{GeV}$ (FWHM energy spread $\SI{1.3}{GeV}$, bunch charge $\SI{6}{pC}$).

Maximal wake-to-bunch energy transfer also requires beam loading of the bunch current \cite{1987beam, Esarey.2009}. For bunch currents with strong beamloading, the overall acceleration gradient is reduced. Increasing the dopant concentration to $\SI{5}{\%}$ evidenced this [see figure \ref{fig:beams}(d)]. The estimated total charge increased by a factor of $\gtrsim 3$, but resulted in a maximum bunch energy of $\SI{7.44}{GeV}$ ($\SI{0.25}{GeV}$ FWHM energy spread). 

We note that future, high repetition-rate laser systems (e.g., those based on fiber lasers \cite{jauregui2013high}) allow for precise control over the transverse laser modes as opposed to bulk-crystal flat-top-like beams currently available. For the same laser energy and channel density, although with the pulse length optimized for electron beam trapping ($\approx \SI{70}{fs}$) \cite{Chen2012}, the orange curve in figure \ref{fig:spectrum}(c) shows the laser intensity evolution for a linearly-matched input mode. Mode beating was greatly reduced, but not eliminated because at these intensities the plasma refractive index itself is modified, slice-by-slice along the pulse, by relativistic self-focusing and ponderomotive self-channeling \cite{Sprangle1990a, sprangle1992interaction, Benedetti2012}. This simulation resulted in the production of a $\SI{13.0}{GeV}$, $\SI{65}{pC}$ electron bunch, trapped in the region $z \lesssim \SI{4}{cm}$, after a $\SI{50}{cm}$ propagation in plasma. This was then compared to the green curve of figure \ref{fig:spectrum}(c) where the transverse laser mode on each longitudinal slice of the pulse was varied so that it remained matched to the ponderomotively perturbed channel (super-matching) \cite{Benedetti2015}. Mode beating, and oscillations in $\hat{a}(z)$ were completely eliminated, permitting the wake structure to remain constant throughout. The simulated electron bunch was  $\SI{13.1}{GeV}$, $\SI{102}{pC}$ after a $z \approx \SI{50}{cm}$. This demonstrates clearly that control over the laser mode maximizes coupling efficiency to the $(0,0)$ channel mode, and minimizes the distance over which over which the wakefield is unsuitable for electron beam transport. 
The marginal difference in energy gain between a laser pulses initiated with a super-matched or a linearly-matched input mode is critically encouraging, since it could mitigate the necessity for super-matching. 
We also note that tailoring of the plasma channel at the entrance of the plasma channel \cite{antonsen1995leaky, Kim2002, Picksley2020, Picksley2023} to filter $p,m > 0$ modes more rapidly could also reduce the mode-beating distance in current systems.

In conclusion, unprecedented insight into the mechanisms of laser propagation in meter-scale LPAs has been gained through varying the accelerator length on a shot-by-shot basis for the first time. Using extensive optical diagnostics, we observed laser coupling into high-order channel modes and their energy loss through mode-filtering, followed by quasi-matched propagation of the fundamental mode, and non-linear depletion of laser energy to the wakefield. We quantified the reduction in laser-to-wake efficiency and electron energy gain caused by the laser mode of currently available PW-class laser systems, and showed how precise control over the mode can result in a significant increase of the bunch energy and charge for the same channel. Matched guiding at $n_0 \approx \SI{e17}{cm^{-3}}$ suppressed self-trapping of electrons throughout the structure. With the introduction of nitrogen dopant, electron beams were with single, quasimonoenergetic peaks to $\SI{9.2}{GeV}$ were generated with $\SI{21.3}{J}$ of laser energy. We note that previous demonstrations of acceleration to $\sim \SI{7.8}{GeV}$ using the same laser required energy of $\SI{31}{J}$ and produced electron spectra containing several lower energy peaks \cite{Gonsalves2019}. This work opens the door for advanced injection techniques to trap ultra-low emittance bunches \cite{Yu2014} using plasma structures well-suited to repetition rates exceeding $\SI{1}{kHz}$ \cite{Shalloo2018, Alejo2022}, meeting vital requirements for future compact accelerators. 

\begin{acknowledgments}
This work was supported by the Director, Office of Science, Office of High Energy Physics, of the U.S. Department of Energy under Contract No. DE-AC02-05CH11231, the Defense Advanced Research Projects Agency, and used the computational facilities at the National Energy Research Scientific Computing Center (NERSC). E. Rockafellow is supported by NSF GRFP (DGE 1840340). We greatly acknowledge technical support from Zac Eisentraut, Mark Kirkpatrick, Federico Mazzini, Nathan Ybarrolaza, Derrick McGrew, Teo Maldonado Mancuso, Art Magana and Joe Riley. The authors would like to thank Nathan Cook,  Jens Osterhoff, Davide Terzani, Remi Lehe, Liona Fan-Chiang, Lieselotte Obst-Huebl, Marlene Turner, and Aodhan McIlvenny for useful discussions. We thank Samantha Trieu (LBNL Creative Services) and Chetanya Jain for design support in Figure \ref{fig:setup}.
\end{acknowledgments}
\newpage

\bibliography{references}

\begin{thebibliography}{80}
\bibitem [{sup()}]{superGauss}%
  \BibitemOpen
  \href@noop {} {}\bibinfo {note} {{A super-Gaussian laser intensity profile is
  defined as $I = A \exp[-(x/w)^n]$ where $n$ is the order of the fit; $n = 2$
  represents a Gaussian, and $n \gg 1$ indicates a top-hat profile
  }}\BibitemShut {NoStop}%
\end{thebibliography}

\end{document}